\input harvmac
\let\includefigures=\iftrue
\let\useblackboard=\iftrue
\newfam\black

\includefigures
\message{If you do not have epsf.tex (to include figures),}
\message{change the option at the top of the tex file.}
\input epsf
\def\figin{\epsfcheck\figin}\def\figins{\epsfcheck\figins}
\def\epsfcheck{\ifx\epsfbox\UnDeFiNeD
\message{(NO epsf.tex, FIGURES WILL BE IGNORED)}
\gdef\figin##1{\vskip2in}\gdef\figins##1{\hskip.5in}
\else\message{(FIGURES WILL BE INCLUDED)}%
\gdef\figin##1{##1}\gdef\figins##1{##1}\fi}
\def\DefWarn#1{}
\def\figinsert{\goodbreak\midinsert}
\def\ifig#1#2#3{\DefWarn#1\xdef#1{fig.~\the\figno}
\writedef{#1\leftbracket fig.\noexpand~\the\figno}%
\figinsert\figin{\centerline{#3}}\medskip\centerline{\vbox{
\baselineskip12pt\advance\hsize by -1truein
\noindent\footnotefont{\bf Fig.~\the\figno:} #2}}
\endinsert\global\advance\figno by1}
\else
\def\ifig#1#2#3{\xdef#1{fig.~\the\figno}
\writedef{#1\leftbracket fig.\noexpand~\the\figno}%
\global\advance\figno by1} \fi

\def\id{{1 \kern-.28em {\rm l}}}
\def\N{{\cal N}}
\def\O{{\cal O}}

\def\K3{{\bf K3}}
\def\journal#1&#2(#3){\unskip, \sl #1\ \bf #2 \rm(19#3) }
\def\andjournal#1&#2(#3){\sl #1~\bf #2 \rm (19#3) }

\def\bar{\overline}
\def\hat{\widehat}
\def\ie{{\it i.e.}}
\def\eg{{\it e.g.}}

\def\tilde{\widetilde}

\def\frac#1#2{{#1\over#2}}

\def\half{\frac12}

\def\inbar{\,\vrule height1.5ex width.4pt depth0pt}
\def\IC{\relax\hbox{$\inbar\kern-.3em{\rm C}$}}
\def\IR{\relax{\rm I\kern-.18em R}}
\def\IP{\relax{\rm I\kern-.18em P}}

%
%

%
\catcode`\@=11
\def\slash#1{\mathord{\mathpalette\c@ncel{#1}}}
\overfullrule=0pt

\def\PP{{\cal P}}

\def\underrel#1\over#2{\mathrel{\mathop{\kern\z@#1}\limits_{#2}}}

\catcode`\@=12


%

\def\O{{\cal O}}
\def\P{{\cal P}}
\def\N{{\cal N}}
\def\ie{{\it i.e.}}
\def\eg{{\it e.g.}}


\lref\KutasovIY{
  D.~Kutasov, A.~Parnachev and D.~A.~Sahakyan,
  ``Central charges and U(1)R symmetries in N = 1 super Yang-Mills,''
  JHEP {\bf 0311}, 013 (2003)
  [arXiv:hep-th/0308071].
}

\lref\IntriligatorMI{
  K.~A.~Intriligator and B.~Wecht,
  ``RG fixed points and flows in SQCD with adjoints,''
  Nucl.\ Phys.\  B {\bf 677}, 223 (2004)
  [arXiv:hep-th/0309201].
}

\lref\GreenDA{
  D.~Green, Z.~Komargodski, N.~Seiberg, Y.~Tachikawa and B.~Wecht,
  ``Exactly Marginal Deformations and Global Symmetries,''
  arXiv:1005.3546 [hep-th].
}

\lref\IntriligatorJJ{
  K.~A.~Intriligator and B.~Wecht,
  ``The exact superconformal R-symmetry maximizes a,''
  Nucl.\ Phys.\  B {\bf 667}, 183 (2003)
  [arXiv:hep-th/0304128].
}

\lref\BarnesJJ{
  E.~Barnes, K.~A.~Intriligator, B.~Wecht and J.~Wright,
  ``Evidence for the strongest version of the 4d a-theorem, via  a-maximization
  along RG flows,''
  Nucl.\ Phys.\  B {\bf 702}, 131 (2004)
  [arXiv:hep-th/0408156].
}

\lref\BarnesZN{
  E.~Barnes, K.~A.~Intriligator, B.~Wecht and J.~Wright,
  ``N = 1 RG flows, product groups, and a-maximization,''
  Nucl.\ Phys.\  B {\bf 716}, 33 (2005)
  [arXiv:hep-th/0502049].
}

\lref\CsakiUJ{
  C.~Csaki, P.~Meade and J.~Terning,
  ``A mixed phase of SUSY gauge theories from a-maximization,''
  JHEP {\bf 0404}, 040 (2004)
  [arXiv:hep-th/0403062].
}

\lref\OkudaME{
  T.~Okuda and Y.~Ookouchi,
  ``Higgsing and superpotential deformations of ADE superconformal  theories,''
  Nucl.\ Phys.\  B {\bf 733}, 59 (2006)
  [arXiv:hep-th/0508189].
}

\lref\KutasovXU{
  D.~Kutasov and A.~Schwimmer,
  ``Lagrange multipliers and couplings in supersymmetric field theory,''
  Nucl.\ Phys.\  B {\bf 702}, 369 (2004)
  [arXiv:hep-th/0409029].
}

\lref\KutasovUX{
  D.~Kutasov,
  ``New results on the 'a-theorem' in four dimensional supersymmetric field
  theory,''
  arXiv:hep-th/0312098.
}

\lref\LeighEP{
  R.~G.~Leigh and M.~J.~Strassler,
  ``Exactly Marginal Operators And Duality In Four-Dimensional N=1
  Supersymmetric Gauge Theory,''
  Nucl.\ Phys.\  B {\bf 447}, 95 (1995)
  [arXiv:hep-th/9503121].
}

\lref\AharonyHX{
  O.~Aharony, B.~Kol and S.~Yankielowicz,
  ``On exactly marginal deformations of N = 4 SYM and type IIB  supergravity on
  JHEP {\bf 0206}, 039 (2002)
  [arXiv:hep-th/0205090].
}

\lref\NovikovUC{
V.~A.~Novikov, M.~A.~Shifman, A.~I.~Vainshtein and V.~I.~Zakharov,
``Exact Gell-Mann-Low Function Of Supersymmetric Yang-Mills Theories From Instanton Calculus,''
Nucl.\ Phys.\ B {\bf 229}, 381 (1983).
}

\lref\ShifmanZI{
M.~A.~Shifman and A.~I.~Vainshtein,
``Solution Of The Anomaly Puzzle In Susy Gauge Theories And The Wilson Operator Expansion,''
Nucl.\ Phys.\ B {\bf 277}, 456 (1986)
[Sov.\ Phys.\ JETP {\bf 64}, 428 (1986\ ZETFA,91,723-744.1986)].
}

\lref\BanksNN{
  T.~Banks and A.~Zaks,
  ``On The Phase Structure Of Vector-Like Gauge Theories With Massless
  Fermions,''
  Nucl.\ Phys.\  B {\bf 196}, 189 (1982).
}

\lref\SeibergPQ{
  N.~Seiberg,
  ``Electric - magnetic duality in supersymmetric nonAbelian gauge theories,''
  Nucl.\ Phys.\  B {\bf 435}, 129 (1995)
  [arXiv:hep-th/9411149].
}

\lref\KolZT{
  B.~Kol,
  ``On conformal deformations,''
  JHEP {\bf 0209}, 046 (2002)
  [arXiv:hep-th/0205141].
}

\lref\KolUB{
  B.~Kol,
  ``On Conformal Deformations II,''
  arXiv:1005.4408 [hep-th].
}

\lref\ZamolodchikovTI{
  A.~B.~Zamolodchikov,
  ``Renormalization Group and Perturbation Theory Near Fixed Points in
  Two-Dimensional Field Theory,''
  Sov.\ J.\ Nucl.\ Phys.\  {\bf 46}, 1090 (1987)
  [Yad.\ Fiz.\  {\bf 46}, 1819 (1987)].
}

\lref\ShapereUN{
  A.~D.~Shapere and Y.~Tachikawa,
  ``A counterexample to the 'a-theorem',''
  JHEP {\bf 0812}, 020 (2008)
  [arXiv:0809.3238 [hep-th]].
}

\lref\KutasovVE{
  D.~Kutasov,
  ``A Comment on duality in N=1 supersymmetric nonAbelian gauge theories,''
  Phys.\ Lett.\  B {\bf 351}, 230 (1995)
  [arXiv:hep-th/9503086].
}

\lref\KutasovNP{
  D.~Kutasov and A.~Schwimmer,
  ``On duality in supersymmetric Yang-Mills theory,''
  Phys.\ Lett.\  B {\bf 354}, 315 (1995)
  [arXiv:hep-th/9505004].
}

\lref\KutasovSS{
  D.~Kutasov, A.~Schwimmer and N.~Seiberg,
  ``Chiral Rings, Singularity Theory and Electric-Magnetic Duality,''
  Nucl.\ Phys.\  B {\bf 459}, 455 (1996)
  [arXiv:hep-th/9510222].
}

\Title{} {\centerline{$a$-Maximization, Global Symmetries and RG Flows}}

\bigskip
\centerline{\it Denis Erkal and David Kutasov}
\bigskip

\centerline{EFI and Department of Physics, University of
Chicago}\centerline{5640 S. Ellis Av. Chicago, IL 60637}

\smallskip

\vglue .3cm

\bigskip

\bigskip
\noindent
We construct the $a$-function of \KutasovUX\ for general F-term 
perturbations of a $3+1$ dimensional $\N=1$ SCFT. We use this 
construction to argue that the central charge $a$ always decreases 
along the corresponding RG flows, and discuss some other applications.

\bigskip

\Date{July 2010}


\newsec{Introduction}

$a$-maximization was originally proposed in \IntriligatorJJ\ as a
way to determine the $U(1)_R$ subgroup of the global symmetry group
of a four dimensional $\N=1$ superconformal field theory (SCFT) which
belongs to the superconformal multiplet. Since the charge of a chiral
operator $\O$ under the superconformal $U(1)_R$, $R(\O)$, is related
to its dimension, $\Delta(\O)$, by the superconformal algebra,
\eqn\qdelta{\Delta(\O)={3\over2} R(\O)~,}
determining the $U(1)_R$ charges is tantamount to determining
the scaling dimensions of chiral operators, which is clearly of
interest, especially in interacting SCFT's.

The authors of \IntriligatorJJ\ showed that the superconformal $U(1)_R$ satisfies the
following property. The central charge $a$, given by\foot{Conventionally, $a$ is defined with
an additional factor $3/32$, which we will omit here.}
\eqn\adefintro{ a = 3 {\rm tr} R^3 - {\rm tr} R ~,}
which can be evaluated for any $U(1)_R$ subgroup of the global symmetry group,
is locally maximized for the superconformal $U(1)_R$. The quantities ${\rm tr} R^n$ in
\adefintro\ stand for the cubic anomaly of three $R$ currents (for $n=3$), or one $R$
current and two stress tensors (for $n=1$). The content of \adefintro\ is to indicate
the absolute normalization of these anomalies -- for a free field theory of a chiral
superfield $\Phi$ with R-charge $R_\phi$ (so that the R-charge of the fermion in the
multiplet is $R_\psi=R_\phi-1$), ${\rm tr} R^n=R_\psi^n$.

Many non-trivial SCFT's  can be obtained as infrared limits of asymptotically
free gauge theories, possibly with non-zero superpotential. The results of \IntriligatorJJ\
(extended in \KutasovIY\ to account for a certain class of accidental symmetries) allow
one to determine the scaling dimensions of chiral operators in such theories (see \eg\
\refs{\IntriligatorMI\KutasovUX\CsakiUJ\BarnesJJ-\OkudaME}).

In \KutasovUX, one of us proposed a  generalization of $a$-maximization away from fixed
points of the renormalization group (RG). In that paper it was shown that one can define
a function $a(\lambda_\alpha)$ on a space of certain interpolating parameters $\lambda_\alpha$,
which coincides with the central charge $a$ obtained from  $a$-maximization
at fixed points of the RG. The interpolating function $a(\lambda_\alpha)$ was used
to show that under some assumptions, the central charge of the IR fixed point
is always smaller than that of the UV fixed point, $a_{IR}<a_{UV}$.

It was further proposed in \KutasovUX\ that the interpolating
parameters $\lambda_\alpha$ provide a parametrization of the space
of couplings of the corresponding quantum field theory. Thus, one
can think of $a(\lambda_\alpha)$  as a generalization of the central
charge $a$ to non-conformal theories. Evidence for this claim was
provided by matching results for anomalous dimensions in weakly
coupled gauge theory with the analysis coming from $a$-maximization.
Additional discussion of this correspondence appeared
in \refs{\BarnesJJ,\KutasovXU,\BarnesZN}.

In the formalism of \KutasovUX, RG flow is a gradient flow,
with $a(\lambda_\alpha)$ playing the role of the potential
\refs{\KutasovUX,\BarnesJJ,\KutasovXU}. The $\beta$-functions
of the couplings are related to the gradient of $a$,
\eqn\gradflow{\partial_\alpha a={\hat
G}_{\alpha\gamma}\beta^\gamma~,} where $\hat G$ is a positive definite
metric on coupling space.
Thus, $a$ decreases under relevant perturbations, increases under
irrelevant ones, and remains constant on moduli spaces of SCFT's.

The papers described above focused on deformations of free field theories by a combination
of gauge interactions and superpotentials for the chiral superfields. Our main purpose in
this note is to generalize the discussion to perturbations of a general $\N=1$ SCFT, $\P$,
by a superpotential\foot{Gauge interactions will also be included, as in \KutasovUX.}
\eqn\Wsuper{ W = \sum_\alpha g^\alpha  \O_\alpha~,}
where $\O_\alpha$ are (local) chiral operators in $\P$, and
$g^\alpha$ are the corresponding coupling constants.  In some
cases, such deformations can be treated using the results of
\KutasovUX. We will see that thinking about the general
case leads to new insights into supersymmetric RG flows,
$a$-maximization and  the role of accidental symmetries. We will
also make contact with the recent discussion of marginal deformations
of $\N=1$ SCFT's \GreenDA.

We start in section 2 with a brief review of some of the results of \KutasovUX.
In section 3 we generalize them to superpotential deformations of a SCFT which
is assumed to have a set of global symmetries acting in a particular way on the
deformations $\O_\alpha$ \Wsuper. In section 4 we present our generalization
of the construction of \KutasovUX\ to deformations of an arbitrary SCFT
corresponding to a combination of gauging part of the global symmetry group and
turning on a superpotential \Wsuper. We discuss it further in section 5.


\newsec{Review of \KutasovUX}

Consider a supersymmetric gauge theory with gauge group $G$ and
chiral superfields $\Phi_i$ in the representations $r_i$ of the
gauge group.\foot{We will take the gauge group to be simple for
concreteness. It is easy to generalize the discussion to product
gauge groups.} One can choose a basis of generators of the gauge
group in the representation $r$, $T^a$, such that
\eqn\ttrr{{\rm tr}_r(T^aT^b)=T(r)\delta^{ab}~.}
The invariant \ttrr\ for the
adjoint representation will be denoted by $T(G)$. The NSVZ
$\beta$-function for the gauge coupling $\alpha=g^2/4\pi$ is
\refs{\NovikovUC,\ShifmanZI}
\eqn\betal{\beta(\alpha)=-{\alpha^2\over2\pi}{3T(G)-\sum_i
T(r_i)(1-\gamma_i(\alpha))\over 1-{\alpha\over2\pi}T(G)}~.}
Here $\gamma_i$ is the anomalous dimension of $\Phi_i$; at weak
coupling one has
\eqn\weakgamma{\gamma_i(\alpha)=-{\alpha\over\pi} C_2(r_i)+
O(\alpha^2)~,}
where $C_2(r)$ is the quadratic Casimir in the representation $r$,
$T^aT^a=C_2(r)I$.

At fixed points
of the RG, $\alpha=\alpha^*$, the scaling dimension of $\Phi_i$ is
given by \eqn\dimgam{\Delta(\Phi_i)=1+{1\over2}\gamma_i(\alpha^*)~.}
Using \qdelta, one can relate the anomalous dimension to the
corresponding R-charge,
\eqn\rig{R_i={2+\gamma_i\over3}~.}
Since $\gamma_i(\alpha)\to 0$ as $\alpha\to 0$ (see \weakgamma),
eq. \betal\ implies that $\alpha$ is marginally relevant (\ie\ the theory
is asymptotically free) when $3T(G)-\sum_iT(r_i)>0$. In this case
$\alpha$ grows at long distances, and at the IR fixed point one has
\eqn\irfixed{3T(G)-\sum_i T(r_i)(1-\gamma_i(\alpha^*))=0~,}
where we assumed that the coupling
$\alpha^*$ is sufficiently small that the denominator in \betal\
remains regular all the way from the UV to the IR.\foot{This can be
shown to be the case when $3T(G)-\sum_iT(r_i)$ is below a certain
critical value.} By using \rig, one can rewrite \irfixed\ as a
condition on the superconformal R-charges of $\Phi_i$ at the IR
fixed point, $R_i(\alpha^*)$, \eqn\anomcond{T(G)+\sum_i
T(r_i)\left[R_i(\alpha^*)-1\right]=0~.} This condition can be
alternatively interpreted as the requirement that the R-symmetry
under which the chiral superfields $\Phi$ have charges
$R_i(\alpha^*)$ (and the super-coordinate $\theta$ has charge one)
is anomaly-free, and thus conserved throughout the RG flow.

A useful way of thinking about the origin of the condition \anomcond\
is the following. The UV fixed point of the gauge theory is a free
field theory, for which $\alpha$ and the anomalous dimensions $\gamma_i$
vanish, and the superconformal R-charges \rig\ are $R_i=2/3$. In addition
to this R-symmetry, this theory has a set of non-R global $U(1)$ symmetries,
one for each irreducible representation $r_i$ (the full global symmetry
group may be larger, but this is unimportant for our purposes). For
non-zero gauge coupling, one combination of the superconformal $U(1)_R$
and the other $U(1)$'s is broken by the chiral anomaly.

A particular combination of the corresponding currents belongs to
the supercurrent multiplet (which also includes the stress-tensor
and the supersymmetry current). One can think of this current
as that under which the superfields $\Phi_i$ have charges $R_i(\alpha)$,
which are related to the running anomalous dimensions $\gamma_i(\alpha)$
via \rig. This  combination is in general not conserved, due to the
non-vanishing $\beta$-function. However, at the fixed points
$\alpha=0,\alpha^*$, where the theory becomes conformal, it flows to
the superconformal $U(1)_R$ symmetry, which must be conserved.

If we assume that the superconformal $U(1)_R$ symmetry at the IR fixed
point, $\alpha=\alpha^*$, is visible in the UV (as opposed to being an
accidental symmetry of the IR fixed point, a possibility we will return
to below), it must be anomaly free, \anomcond. This assumption is
plausible since by choosing the gauge group $G$ and representations $r_i$
appropriately, one can arrange for the IR fixed point to be at arbitrarily
weak coupling \BanksNN; in that case, accidental symmetries are impossible.

The procedure proposed in \KutasovUX\ for determining $R_i(\alpha)$
was the following. Consider the function \eqn\genera{
a(R_i,\lambda_G)= 2| G|+\sum_i |r_i|\left[
3(R_i-1)^3-(R_i-1)\right]- \lambda_G\left[T(G)+\sum_i
T(r_i)(R_i-1)\right]~,} where $|G|$ is the dimension of the gauge
group, and $|r|$, the dimension of the representation $r$. One can
think of the first two terms in \genera\ as the central charge
\adefintro\  evaluated for the symmetry which assigns R-charge $R_i$
to $\Phi_i$. Varying the last term with respect to the Lagrange
multiplier $\lambda_G$ enforces the constraint \anomcond\ which
characterizes the non-trivial fixed point of the SYM theory. Thus,
$\lambda_G$ plays here a role analogous to that of the gauge
coupling in standard QFT discussions. For $\lambda_G=0$, \genera\ is
the central charge of the free field theory corresponding to the UV
fixed point $(\alpha=0)$; indeed, locally maximizing \genera\ w.r.t.
$R_i$ gives the UV values ($R_i=2/3$) \IntriligatorJJ. On the other
hand, as shown in \KutasovUX, if we vary \genera\ w.r.t. $R_i$ at
fixed $\lambda_G$, find the local maximum $R_i(\lambda_G)$, and then
extremize the resulting $a(R_i(\lambda_G),\lambda_G)$ w.r.t.
$\lambda_G$, we get the correct central charge at the IR fixed
point, $\alpha=\alpha^*$.

It is thus very natural to think of $R_i(\lambda_G)$  and
$a(R_i(\lambda_G), \lambda_G)$ as the scale dependent R-charges and
central charge along the RG flow from the UV to the IR. As we will
see below, this procedure can be thought of as determining the
$U(1)$ that belongs to the current multiplet along the RG flow. The
anomalous dimensions $\gamma_i(\lambda_G)$, which come from
substituting $R_i(\lambda_G)$ into \rig, have been compared with
perturbative results in \refs{\BarnesJJ,\KutasovXU} and found to
agree with them up to three loops (the order to which they have been
calculated perturbatively).

At weak coupling, the Lagrange multiplier $\lambda_G$ is related to
the gauge coupling via the relation
\eqn\rellamal{\lambda_G={2\alpha\over\pi}|G|+O(\alpha^2)~.}
Thus, $\lambda_G$ is proportional to the 't Hooft coupling of the gauge
theory. The higher order terms in \rellamal\ suffer from well known
ambiguities associated with reparametrizations of the space of
couplings. Since we do not know a priori how the parametrization
employed in \KutasovUX\ is related to that corresponding to a
particular perturbative calculation, the comparisons in
\refs{\BarnesJJ,\KutasovXU} have some freedom, which was
taken into account in these papers.

The above discussion is valid when the gauge coupling $\alpha$ remains sufficiently weak
throughout the RG flow, that we do not have to worry about the denominator of the NSVZ
$\beta$-function \betal. It is not understood in general what constraint this places on
the gauge group and matter representations, but it is known that there are cases in which
this assumption breaks down; a famous example \SeibergPQ\ is supersymmetric QCD with
$N_f<{3\over2}N_c$. In such cases, new symmetries appear along the RG flow, and they need
to be included in the analysis. We will comment on the role of such accidental symmetries below.

So far, we reviewed the case of free field theories perturbed by gauge interactions. One can also
turn on a (gauge-invariant) superpotential $W(\Phi_i)$ for the chiral superfields $\Phi_i$. This
was also discussed in \KutasovUX, but we will not review the details here. Instead, in the next 
section we will take the first step in generalizing the discussion of \KutasovUX\ to perturbations of
more general fixed points. This will allow us to present the basic ideas of \KutasovUX\ while
introducing some elements that will be useful for the general discussion in section 4.


\newsec{Superpotential perturbations and diagonal abelian symmetries}

We start with an SCFT, $\P$, whose superconformal $U(1)_R$ will be
denoted by $J^\mu_\P$. We are interested in perturbing the theory as
in \Wsuper, and studying how the $U(1)_R$ that belongs to the current 
multiplet is deformed by the perturbation. In order for such a deformation
to be possible, the theory $\P$ must have some additional 
(non-R) symmetries;\foot{The deformation of $J^\mu_\P$ must be
by the current component of a real superfield whose dimension is
two  and is thus conserved. Real superfields with dimension larger
than two cannot contribute to the deformation in a finite region around
$\P$.} we will denote the corresponding conserved currents by $J^\mu_a$. 
In this section, we will make two simplifying assumptions about them:
\item{(1)} We will take the global symmetry group to be $U(1)^n$.
For our discussion, only symmetries that act non-trivially on the
deformations \Wsuper\ play a role. Thus, theories with non-abelian
symmetries that commute with the perturbations are included in the
analysis of this section.
\item{(2)} We will assume that the charge operators $Q_\P$ and $Q_a$
corresponding to $J_\P^\mu$ and $J_a^\mu$ act diagonally on $\O_\alpha$,
\eqn\AbelComm{ \eqalign{Q_\P\O_\alpha&= R_\P^\alpha \O_\alpha~,\cr
Q_a\O_\alpha&= q_a^\alpha\O_\alpha~.} }

\noindent
Both conditions will be relaxed in the next section, where we will also
discuss the generalization of the construction to include gauge interactions.

In the presence of the perturbation \Wsuper, the $U(1)_R$ current in the
supercurrent multiplet, $J^\mu_g$, is in general deformed, and becomes a
linear combination of $J^\mu_\P$ and $J^\mu_a$,
\eqn\JgAbel{ J^\mu_g = J^\mu_\P + d^a J^\mu_a ~,}
where the $d^a$ are functions of the couplings $g^\alpha$ (which vanish
for $g^\alpha=0$). The charge of $\O_\alpha$ under $Q_g=Q_\P+d^a Q_a$ 
can be read off from \AbelComm:
\eqn\Rperturb{ R^\alpha = R^\alpha_\P + d^a q^\alpha_a~. }
Interpreting $R^\alpha$ as the running R-charge along the RG flow, as
before, the $\beta$-function for the coupling $g^\alpha$ \Wsuper\ is
(see \eg\ \LeighEP)
\eqn\betaf{\beta^\alpha=-{3\over2}g^\alpha(2-R^\alpha)~.}
At non-trivial fixed points of the RG the perturbations $\O_\alpha$ must be
marginal, \ie\ the R-charges \Rperturb\ must approach $2$. It is thus natural,
following the logic of \KutasovUX, to introduce the interpolating ``$a$-function''
\eqn\aAbeldef{ a(\lambda, d) = \left(3 {\rm tr} Q_g^3 - {\rm tr} Q_g\right) -
\sum_\alpha \lambda_\alpha(2-R^\alpha)~. }
The first term in \aAbeldef\ is the central charge $a$ for the symmetry
\JgAbel\ in the SCFT $\P$. In the second term, the Lagrange multipliers
$\lambda_\alpha$ impose the condition for a fixed point $R^\alpha=2$.

To evaluate \aAbeldef, we plug \JgAbel\ into it. The first term
gives a cubic polynomial in $d^a$,
\eqn\cubpol{3 {\rm tr} Q_g^3 - {\rm tr}
Q_g=3D_{abc}d^ad^bd^c-3\tau_{ab}d^ad^b+a(\P)~.}
$D_{abc}$ is the triangle anomaly of the global currents,
\eqn\dabc{D_{abc}={\rm tr} Q_aQ_bQ_c~,}
and $\tau_{ab}$ is their two point function, evaluated in the SCFT $\P$,
\eqn\JmuJnu{ \langle J^\mu_a (x) J^\nu_b (0) \rangle =
{\tau_{ab}\over(2\pi)^4} (\partial^\rho
\partial_\rho \eta^{\mu \nu} - \partial^\mu \partial^\nu)
\frac{1}{|x|^4} ~.}
Unitarity of $\P$ implies that the matrix $\tau_{ab}$ is positive definite.
To arrive at \cubpol\ we used the results of \IntriligatorJJ,
\eqn\TrJRR{\eqalign{9 {\rm tr} Q_{\cal P}^2 Q_a =& {\rm tr} Q_a~;\cr
{\rm tr} Q_\P Q_a Q_b =& - \frac{\tau_{ab}}{3}~.\cr}}
In particular, the absence of a term linear in $d^a$ in \cubpol, and the fact
that $d^a=0$ is a local maximum of this expression, are consequences
of the original $a$-maximization \IntriligatorJJ, applied to $\P$.

Adding the last term in \aAbeldef, we find\foot{Sums over $a$,
$\alpha$ are implied.}
\eqn\aAbela{ a(\lambda,d) = a({\cal P}) - 3 \tau_{ab}d^ad^b
+ 3D_{abc}d^a d^b d^c  + \lambda_\alpha q^\alpha_a d^a
- \lambda_\alpha (2-R_{\cal P}^\alpha)~.}
Following \KutasovUX\ we interpret $\lambda_\alpha$ as coupling constants,
which are related to $g^\alpha$ in \Wsuper\  (the precise relation
will be presented below). The parameters $d^a$, like the $R_i$ in
the previous section, are determined by locally maximizing \aAbela,
\eqn\aAbelmax{ \frac{\partial a}{\partial d^a} = - 6
\tau_{ab} d^b + 9 D_{abc} d^b
d^c +  \lambda_\alpha q^\alpha_a=0~.}
Solving the quadratic equation for $d^a(\lambda)$ and substituting
back into \aAbela\ leads to
\eqn\alambda{ a(\lambda) = a(\P) + 3 \tau_{ab}
d^a d^b - 6 D_{abc} d^a
d^bd^c - \lambda_\alpha (2-R^\alpha_{\P})~.}
Equations \aAbelmax, \alambda\ describe the full dependence of the generalized central charge
$a$ on the couplings $\lambda_\alpha$ in terms of data associated with the SCFT $\P$.

For small $\lambda_\alpha$, we can approximate the solution of \aAbelmax\
as\foot{Note that we used the metric $\tau_{ab}$ to raise the index on $q_a^\alpha$.}
\eqn\cAbel{d^a =\frac{1}{6} \lambda_\alpha q^{a\alpha } +O(\lambda^2)~.}
The R-charges \Rperturb\ are given by
\eqn\rsmall{R^\alpha=R^\alpha_\P+{1\over6} q_a^\alpha
q^{a\beta}\lambda_\beta+O(\lambda^2)~.}
Since the operators $\O_\alpha$ are terms in a superpotential \Wsuper,
it is natural to define their anomalous dimensions $\gamma^\alpha$ by
\eqn\anomdims{\Delta(\O_\alpha)=3+\half\gamma^\alpha~.} Using
\qdelta, \rsmall\ we find \eqn\anomsmall{\gamma^\alpha=3(R^\alpha-2)
= 3(R^\alpha_\P-2) + \half q_a^\alpha
q^{a\beta}\lambda_\beta+O(\lambda^2)~.}
Plugging \cAbel\ into \alambda, we find
\eqn\aAbelb{ a(\lambda) =
a(\P)-(2-R^\alpha_\P)\lambda_\alpha+ \frac{1}{12}
q_a^{\alpha} q^{a\beta} \lambda_\alpha\lambda_\beta
+{\cal O}(\lambda^3)~. }
As expected, we see that relevant perturbations,
corresponding to couplings $\lambda_\alpha$ with $R^\alpha_\P<2$,
decrease the central charge, while irrelevant ones increase it. This
is clear for small $\lambda_\alpha$, but in fact is true exactly.
Indeed, differentiating \alambda\ w.r.t. $\lambda_\alpha$, we find
\eqn\diffal{{da\over d\lambda_\alpha}=-(2-R^\alpha)~,}
where $R^\alpha$ is given by \Rperturb, and $d^a(\lambda)$ are the
solutions of \aAbelmax. Thus, as $\lambda^\alpha$ increases, $a$
decreases (if the perturbation is relevant), and vice-versa.

An interesting special case of \alambda\ is when all the
operators $\O_\alpha$ in \Wsuper\ are marginal at $\P$, \ie\
$R^\alpha_\P=2$ for all $\alpha$. In that case $a(\lambda)$ takes
the form
\eqn\margcase{a(\lambda) = a(\P) + 3 \tau_{ab} d^a d^b - 6
D_{abc} d^ad^bd^c ~.}
We see that in this case the deformed central
charge $a$ depends on the couplings $\lambda^\alpha$ only via the
$d^a$, which solve \aAbelmax. A question that was recently
discussed in \GreenDA\ (see \refs{\LeighEP\AharonyHX\KolZT-\KolUB}
for related discussions), is what are the conditions for marginal
deformations of the sort \Wsuper\ to give rise to a moduli space
of SCFT's. In our formalism, the answer
is clear from inspection of \margcase. Since the central charge $a$
must remain unchanged along such a moduli space, and since the $d^a$
vanish at $\lambda=0$, \cAbel, the necessary and sufficient
condition is $d^a=0$. It is necessary since the metric $\tau_{ab}$
is positive definite, so if any of the $d^a$ turn non-zero, $a(\lambda)$
increases.\foot{Thus, marginal superpotential couplings $g^\alpha$ that
are not exactly marginal are marginally irrelevant, as noted in \GreenDA.}
To see that it is sufficient, note that for $d^a=0$ the operators $\O_\alpha$
remain marginal, since \Rperturb\ implies that $R^\alpha=2$; therefore, the
$\beta$-functions \betaf\ vanish.

Looking back at \aAbelmax, we see that the condition $d^a=0$ implies
that $\lambda_\alpha q^\alpha_a=0$. Thus, truly marginal perturbations
are obtained by taking the coupling vector $\lambda_\alpha$ to be an
eigenvector of the charge matrix $q^\alpha_a$, with eigenvalue zero.
Note that this statement is exact in $\lambda$ (in particular, it does 
not rely on the perturbative expansion \cAbel).

The relation between the couplings  $g^\alpha$ in \Wsuper\ and
Lagrange multipliers $\lambda_\alpha$ in \aAbeldef\ can be fixed by
comparing our result for the anomalous dimensions \anomsmall\ to a
direct perturbative field theory calculation. This calculation was
recently performed in \GreenDA, with the result\foot{The metric
$\gamma_{ab}$ used to raise and lower indices in \GreenDA\ is related
to the one used here via $\tau_{ab}=(2\pi)^4\gamma_{ab}$. Their definition
of the charges $q_a^\alpha$ differs from ours by a factor,
$(q_a^\alpha)_{\rm here}=(2\pi)^2(q_a^\alpha)_{\rm there}$. Finally, it seems
that the $\beta$-function in \GreenDA\ needs to be multiplied by $(2\pi)^2$.}
\eqn\SanomdimAbel{\gamma^\alpha = 8\pi^4 \tau^{ab} q^{\alpha}_a
q^\beta_b g^\beta{\bar g}_\beta+O(|g|^4)~.}
Comparing to \anomsmall\ we find that (no sum over $\alpha$)
\eqn\glamAbel{ \lambda_\alpha = (2\pi)^4 g^\alpha {\bar g}_\alpha+O(|g|^4)~,}
where $\bar g_\alpha={\bar g}^{\bar\beta}h_{\alpha\bar\beta}$
and $h_{\alpha \bar\beta}$ is the metric associated with the two
point function at the conformal point $\cal P$,
\eqn\OOmetric{ \langle {\cal O}_\alpha(x) {\cal O}^{\dag}_{\bar
\beta}(0) \rangle = \frac{h_{\alpha \bar \beta}}{|x|^6} ~.}
The fact that the Lagrange multipliers of \KutasovUX\ are proportional to the
absolute values squared of the couplings in the superpotential was observed
in a special case in \BarnesJJ. We will return to it in the next section.

We saw above that the moduli space of SCFT's is obtained by solving the
constraints $d^a=0$, with $d^a$ the solutions of \aAbelmax. In
\GreenDA\ it was shown that the moduli space satisfies the constraint
$D^a=0$, where $D^a$ are related to the renormalization of terms involving
the currents $J_a$ in the Kahler potential. At weak coupling, comparing the
results of \GreenDA\ with \cAbel, using \glamAbel, one finds that
\eqn\cDAbel{ d^a = {1\over3}D^a~.}
We expect this relation to be exact.

As mentioned above, marginal perturbations with non-zero $d^a$ are in fact
marginally irrelevant. Thus, turning them on corresponds to flowing up the RG
and increasing the generalized central charge $a$. Since the exact behavior of
the anomalous dimensions and of $a$ is known, \Rperturb, \aAbelmax, \margcase,
one can ask whether by following this flow one can reach a non-trivial fixed
point, which would necessarily have
larger $a$. It is easy to see that the answer is negative. Indeed, any non-trivial
fixed point of the RG would have to have vanishing $\beta$-function,
or equivalently $R^\alpha=2$ (see \betaf). Comparing to \Rperturb, this implies
that $d^aq_a^\alpha=0$. On the other hand, multiplying \aAbelmax\ by
$d^a$ and using the above fact, we learn that $2\tau_{ab}d^ad^b=3D_{abc}d^ad^bd^c$.
Plugging this into the expression for $a$, \margcase, we see that at
such a fixed point, $a(\lambda^*)=a(\P)-\tau_{ab}d^ad^b$, which is smaller
(or equal, for $d^a=0$) than the value of $a$ at the IR fixed point $\P$. This
contradicts the fact that the perturbation in question increases $a$.

Hence, we conclude that marginally irrelevant superpotential perturbations of the sort
\Wsuper\ do not lead to non-trivial fixed points. This was observed in the particular
example of the Wess-Zumino model in \KutasovUX, but we see that it is general. Of
course, the above analysis assumes the absence of accidental symmetries, a subject that
we will return to later. For irrelevant perturbations, such symmetries have to do with
the fact that one can embed a particular IR theory in an infinite number of different UV
theories, which from the current perspective differ in their accidental symmetries.

When the perturbing operators $\O_\alpha$ have dimensions slightly below three, \ie\
the R-charges take the form
\eqn\epsexp{R^\alpha_\P=2-\epsilon^\alpha~,}
with $\epsilon^\alpha\ll 1$, one can try to use the $\epsilon$-expansion to find a nearby
IR fixed point. This possibility is familiar from two dimensional QFT \ZamolodchikovTI,
and can occur in four dimensions as well. Substituting \epsexp\ into \rsmall, the condition
for a fixed point, $R^\alpha(\lambda^*)=2$, takes the form
\eqn\ireps{6\epsilon^\alpha=q_a^\alpha q^{a\beta}\lambda_\beta^*+O(\lambda^2)~.}
Thus, the couplings $\lambda_\alpha^*$ are of order $\epsilon$, and one can evaluate
various properties of the IR fixed point by expanding in $\epsilon$.

In \refs{\KutasovUX,\BarnesJJ,\KutasovXU} it was argued that the generalized
central charge $a$ can be thought of as the potential of a gradient flow,
\eqn\betaamax{ \frac{\partial a}{\partial {\bar g}^{\bar \alpha}}
= G_{\bar \alpha\gamma } \beta^\gamma~,}
where $G_{\bar\alpha\gamma }$ is the metric on coupling space. Using \cAbel,
\margcase, \SanomdimAbel, \glamAbel\ we can determine $G_{\bar\alpha\gamma }$
for small $g$:
\eqn\metricG{ G_{\bar\alpha \gamma} = {2\over3}(2\pi)^4
h_{\bar\alpha \gamma}~. }
Note that:
\item{(1)} The relation \metricG\ was derived to leading order in
the couplings, but it is natural to conjecture that it is in fact
valid throughout the RG flow.
\item{(2)} The relation between the Lagrange multipliers and the
couplings \glamAbel\ was obtained for operators $\O_\alpha$ that
are marginal at $\P$. One can show that the same relation, along
with the metric, \metricG, is valid for non-marginal operators.

\noindent
In the analysis above we assumed that when we perturb the SCFT $\P$ by
the superpotential \Wsuper, the R-current mixes only with other symmetry
currents in $\P$ (see \JgAbel). As we mentioned, in some cases new
(``accidental'') symmetries appear along the RG flow, and need to be
included in the discussion. In our framework, this means that for some
values of the couplings, one needs to include additional currents
$J_a$ in the sum \JgAbel. Since our analysis is local in coupling space,
it is in principle easy to add this element to the discussion. One does
not expect it to change the qualitative conclusions; in particular, the
monotonicity of $a$ along RG flows should still be be maintained.


\newsec{The general case}

In this section we extend the discussion of the previous sections to
a general SCFT $\P$ in the presence of general F-term perturbations. We
start by discussing superpotential perturbations \Wsuper, and then
turn to deformations corresponding to gauging a subgroup of the global
symmetry group.

\subsec{Superpotential deformations}

In addition to the superconformal $U(1)_R$ current, $J^\mu_\P$, we assume that
$\P$ has a global symmetry group (in general non-Abelian) with currents $J^\mu_a$.
The perturbing operators $\O_\alpha$ in \Wsuper\ transform under these symmetries
as follows
\eqn\NAbelComm{
\eqalign{Q_\P\O_\alpha &= (R_\P)_\alpha^\beta \O_\beta~,\cr
Q_a \O_\alpha &= (T_a)^\beta_\alpha\O_\beta~,} }
where $R_\P$ is the $U(1)_R$ charge matrix, and $T_a$ are the generators of the
global symmetries $Q_a$. The case of abelian global symmetries acting
diagonally \AbelComm\ corresponds to $(R_\P)_\alpha^\beta=R_\P^\alpha\delta_\alpha^\beta$,
$(T_a)^\beta_\alpha = q^\alpha_a \delta_\alpha^\beta$.

As before, when we turn on a superpotential \Wsuper, the $U(1)_R$ current in
the supercurrent multiplet is deformed to \JgAbel. The $U(1)_R$ charge operator
\eqn\uonech{Q_g=Q_\P+d^aQ_a}
now acts non-diagonally on the $\O_\alpha$:
\eqn\RNAbel{Q_g\O_\alpha=R_\alpha^\beta\O_\beta~,\qquad
R^\beta_\alpha=(R_{\P})_\alpha^\beta + d^a (T_a)^\beta_\alpha ~.}
At first sight, it is not clear how to generalize the construction
of the previous section for this case and, in particular, what is
the analog of the last term in \aAbeldef. To proceed, it is useful
to recall that for a general anomalous dimension matrix, the
$\beta$-functions take the form
\eqn\formbe{\beta^\gamma=-{3\over2}\left(2\delta_\alpha^\gamma-R_\alpha^\gamma\right) g^\alpha~.}
Since we want the generalized central charge $a$ to have the property \betaamax,
it is natural to introduce a complex Lagrange multiplier, $\lambda^\alpha$, and write
\eqn\NAbela{a(\lambda,d) = 3 {\rm tr} Q_g^3 - {\rm tr} Q_g -
\lambda^\alpha {\bar\lambda}_\beta (2\delta^\beta_\alpha - R_\alpha^\beta)~.}
This has the property that the derivative of $a$ w.r.t. ${\bar\lambda}_\beta$
is proportional to the $\beta$-function \formbe\ (assuming that
$\lambda^\alpha\propto g^\alpha$; we will discuss the precise relation between
them below).

We can now proceed as in section 3. Plugging \uonech, \RNAbel\ into \NAbela,
we find
\eqn\NAbelb{ a(\lambda,d) = a(\P) - 3\tau_{ab}d^a d^b +
3 D_{abc}d^a d^b d^c  + d^a \lambda^\alpha {\bar \lambda}_\beta
(T_a)^\beta_\alpha - \lambda^\alpha {\bar \lambda}_\beta
\left[ 2\delta_\alpha^\beta - (R_\P)_\alpha^\beta\right]~.}
Varying w.r.t $d^a$, we find
\eqn\aNAbelmax{ \frac{\partial a}{\partial d^a} = - 6 \tau_{ab} d^b
+ 9 D_{abc} d^b d^c + \lambda^\alpha {\bar \lambda}_\beta
(T_a)^\beta_\alpha = 0~.}
Solving this equation for $d^a(\lambda)$ and plugging back into
\NAbelb\ gives
\eqn\aNAbelc{ a(\lambda) = a(\P) + 3 \tau_{ab} d^a d^b - 6
D_{abc}d^a d^b d^c - \lambda^\alpha {\bar \lambda}_\beta
\left[ 2\delta_\alpha^\beta - (R_\P)_\alpha^\beta\right]~.}
As in section 3, this is the exact result for the generalized central
charge $a$ and running R-charges \RNAbel\ on the space of theories labeled
by the complex parameters $\lambda^\alpha$. To determine the relation
between these parameters and the couplings $g^\alpha$, it is convenient
to expand in $\lambda^\alpha$.
For small $\lambda$, the solution of \aNAbelmax\ can be approximated as
\eqn\cNAbel{ d^a = \frac{1}{6} \lambda^\alpha (T^a)_{\alpha}^{\beta}
{\bar\lambda}_\beta  + {\cal O}(|\lambda|^4) ~.}
Plugging into \RNAbel\ gives
\eqn\rsmallna{R_\alpha^\beta = (R_\P)^\beta_\alpha +
\frac{1}{6} (T_a)^\beta_\alpha \lambda^\gamma
(T^a)_{\gamma}^\delta {\bar\lambda}_\delta +
O(|\lambda|^4)~.}
The resulting anomalous dimension matrix is
\eqn\NAbelgamdef{ \gamma_\alpha^\beta = 3(R_\alpha^\beta -
2\delta_\alpha^\beta) = 3\left[(R_\P)_\alpha^\beta -2\delta_\alpha^\beta\right] +
\frac{1}{2} (T_a)^{\beta}_\alpha\lambda^\gamma
(T^a)_{\gamma}^\delta {\bar\lambda}_\delta +{\cal O}(|\lambda|^4)
~.}
The general result of \GreenDA\ for the marginal case
$(R_\P)_\alpha^\beta=2\delta_\alpha^\beta$ is
\eqn\anomdimNAbel{ \gamma_\alpha^\beta = 8\pi^4
(T_a)^{\beta}_\alpha g^\gamma (T^a)_{\gamma}^\delta
{\bar g}_\delta+O(|g|^4)~.}
Comparing \NAbelgamdef\ and \anomdimNAbel, we see that the Lagrange
multipliers are proportional to the couplings:
\eqn\glamNAbel{ \lambda^\alpha = (2\pi)^2 g^\alpha~,\qquad {\bar
\lambda}_\beta = (2\pi)^2 h_{\beta \bar \beta} {\bar g}^{\bar
\beta}~,}
where $h_{\beta \bar \beta}$ is the metric Zamolodchikov
metric \OOmetric.

Note that the Lagrange multipliers $\lambda^\alpha$ introduced
in section 3 (following \KutasovUX) in the case where the global
symmetry is abelian and acts diagonally, correspond in our present
notation to $\lambda^\alpha\bar\lambda_\alpha$ (as is clear from
comparing \aAbeldef\ and \NAbela). While the Lagrange multipliers
in this section are complex, the corresponding Lagrange multipliers
in section 3 are real and positive, in agreement with the discussion
in \KutasovUX.

The rest of the discussion of section 3
goes through. In particular, the relation between $d^a$
and $D^a$ is the same as there, \cDAbel; the condition
for marginal operators to be truly marginal is still $d^a=0$;
and the gradient flow relation \betaamax\ is still valid,
with the metric given by \metricG. Fixed points of the
RG correspond to stationary points of $a(\lambda,\bar\lambda)$,
and it is easy to see that $a$ decreases under RG flows.

\subsec{Gauging}

Given an SCFT $\P$ with global symmetry $F$, we can gauge
a subgroup\foot{As in section 2, we will take this subgroup
to be simple.} $G$ of $F$. Of course, the superpotential
perturbations \Wsuper\ must be gauge invariant.
The role of the global symmetry in the discussion of the
previous sections is in this case played by $H$, the maximal
subgroup of $F$ that commutes with $G$. In general, part of
$H$ can be anomalous in the gauge theory of $G$, as in section 2.

To construct the generalized central charge $a$ corresponding to the
above gauge theory, we proceed as before. The superconformal $U(1)_R$
current of the SCFT $\P$, $J_\P^\mu$, must be gauge invariant, and
when we turn on the gauge coupling $g$ and superpotential \Wsuper,
it can in general mix with the global currents $J^\mu_a$, $a=1,\cdots,
\dim H$, as in  \JgAbel. The coefficients $d^a$ are functions of the
couplings, $d^a=d^a(g^\alpha, g)$; our goal is to find them.

As discussed in section 2, the last term in equation \genera, which
takes into account the effect of gauge dynamics, is proportional to
the anomaly of the $U(1)_R$ current due to its coupling to the $G$
gauge fields. For the current \JgAbel, this anomaly is given by
\eqn\anomgen{{\rm tr} Q_g Q_A Q_B= {\rm tr} (Q_\P+d^aQ_a) Q_A Q_B\equiv\CA\delta_{AB}~.}
Here $A,B=1,\cdots, \dim G$ run over the gauge group.
To calculate $\CA$ we need to compute the two terms in \anomgen. One has
\eqn\JaABJ{ \eqalign{ {\rm tr} Q_\P Q_A Q_B &= - \frac{\tau_{AB}}{3}=- {k_\P\over3} \delta_{AB}~, \cr
{\rm tr} Q_a Q_A Q_B &= k_a \delta_{AB}
~.}}
On the first line of \JaABJ\ we used \TrJRR\ and the fact that we can choose
a basis of gauge generators for which\foot{Note that positivity of $\tau_{AB}$
implies that $k_\P>0$.}
\eqn\tauu{\tau_{AB}=k_\P\delta_{AB}~.}
On the second line we used the fact that since $G$ and $H$ commute,
the relevant anomaly coefficients are proportional to $\delta_{AB}$.

Plugging \JaABJ\ into \anomgen\ we see that
\eqn\forma{\CA=d^ak_a-{1\over3} k_\P~.}
For example, if $\P$ is the free field theory described in section 2,
$k_\P=\sum_i T(r_i)$, and the second term in \forma\ contributes to
$\CA$ the term $\sum_i T(r_i)(R_i-1)$, with $R_i=2/3$, the free
values. The currents $J_a$ are in this case the global $U(1)$'s
$J_i=\Phi_i^\dagger\Phi_i$, and their contribution to $\CA$ replaces
the free R-charges by arbitrary ones. The full result for the anomaly
coefficient in that case is given by $T(G)+\sum_i T(r_i)(R_i-1)$
(see \anomcond). Thus, \forma\ reproduces everything except the
contribution of the gauginos, $T(G)$, which needs to be added by hand.
This is natural, since \forma\ is the matter contribution, whereas
the gauginos are part of the gauge superfield.

For a general theory, the anomaly of the $U(1)_R$
current is given by $\CA+T(G)$. We can proceed as in the previous sections,
and add the gauge contribution to the generalized central charge \NAbela:
\eqn\agaugeNAB{ a(\lambda,d) = 3 {\rm tr} Q_g^3 - {\rm tr} Q_g -
\lambda^\alpha {\bar \lambda}_\beta (2\delta^\beta_\alpha -
R_\alpha^\beta) - \lambda_G \left(T(G)+ d^a k_a-{1\over3}k_\P\right)~.}
The coefficient of $\lambda_G$ in \agaugeNAB\ is the analog of the
numerator of the NSVZ $\beta$-function \betal\ for  a general theory.
Plugging \cubpol, \RNAbel\ into \agaugeNAB, we find
\eqn\agaugeNABb {\eqalign{ a(\lambda,d) =& a(\P) - 3 \tau_{ab} d^a
d^b + 3 D_{abc} d^a d^b d^c + \cr +&\lambda^\alpha {\bar
\lambda}_\beta d^a (T_a)^\beta_\alpha -
\lambda^\alpha {\bar \lambda}_\beta \left[2\delta^\beta_\alpha - (
R_\P)_\alpha^\beta\right] - \lambda_G \left(T(G)+ d^a
k_a-{1\over3}k_\P\right) ~.}}
Maximizing \agaugeNABb\ w.r.t. $d^a$, we find (compare to \aNAbelmax)
\eqn\Gaugemax{ \frac{\partial a}{\partial d^a} = - 6 \tau_{ab} d^b +
9 D_{abc} d^b d^c + \lambda^\alpha {\bar \lambda}_{\beta}
(T_a)_\alpha^\beta - \lambda_G k_a=0~.}
Solving for $d^a(\lambda)$ and plugging back into \agaugeNABb\ gives
\eqn\agauge{ a(\lambda^\alpha,\bar\lambda_\beta,\lambda_G) = a(\PP) + 3 \tau_{ab} d^a d^b - 6 D_{abc}
d^a d^b d^c - \lambda^\alpha {\bar \lambda}_\beta \left[2\delta^\beta_\alpha - (R_\P)_\alpha^\beta\right] -
\lambda_G \left(T(G) - \frac{1}{3} k_\P\right) ~,}
a generalization of \alambda, \aNAbelc. The coefficient of $\lambda_G$ in \agauge\ is
the one loop $\beta$-function for the $G$ gauge theory. For $T(G)>\frac{1}{3} k_\P$
this gauge theory is asymptotically free, and vice-versa. The analog of the discussion
of marginal perturbations in section 3 is  $(R_\P)_\alpha^\beta=2\delta^\beta_\alpha$,
$T(G)=k_\P/3$. In this case the generalized central charge $a$ takes the form \margcase,
where $d^a$ are obtained by solving \Gaugemax. Truly marginal perturbations correspond
to $d^a=0$, which leads to (see \Gaugemax)
\eqn\trumarg{\lambda^\alpha {\bar \lambda}_{\beta}(T_a)_\alpha^\beta = \lambda_G k_a~.}
This relation between superpotential and gauge couplings is familiar from $\N=4$
SYM, finite $\N=2$ SYM, and related examples.

For small $\lambda_G,\lambda^\alpha$ we can approximate the solution
of \Gaugemax\ as
\eqn\dgauge{ d^a = \frac{1}{6} \Big( \lambda^\alpha
(T^a)_\alpha^\beta {\bar \lambda}_\beta - \lambda_G k^a\Big) +
\cdots~.}
Using this, we can compute the R-charge \RNAbel:
\eqn\Rgauge{R_\alpha^\beta = (R_\P)_\alpha^\beta
+\frac{1}{6} (T_a)^\beta_\alpha \Big( \lambda^\gamma
(T^a)_\gamma^\delta {\bar \lambda}_\delta - \lambda_G k^a \Big)+
\cdots~,}
and anomalous dimension
\eqn\gammagauge{ \gamma_\alpha^\beta = 3(R^\beta_\alpha -
2\delta^\beta_\alpha ) = 3\left[(R_\P)_\alpha^\beta - 2\delta_\alpha^\beta\right] +
\frac{1}{2} (T_a)^\beta_\alpha \Big( \lambda^\gamma
(T^a)_\gamma^\delta {\bar \lambda}_\delta - \lambda_G k^a\Big)+
\cdots~.}
Using the map between the Lagrange multipliers $(\lambda_G, \lambda^\gamma)$,
and the gauge theory couplings $(\alpha,g^\gamma)$, \rellamal, \glamNAbel,
we get a formula for the anomalous dimensions of the gauge theory, which
agrees qualitatively with that of \GreenDA.


\newsec{Discussion}

In this note we extended the construction of the generalized central charge $a$ in
\KutasovUX\ to a general $\N=1$ superconformal field theory $\P$ perturbed by
turning on a superpotential and/or gauging a subgroup of the global symmetry group.
We  emphasized the role of global symmetry currents which can mix with the current
that belongs to the supercurrent multiplet. Our main result is the formula for the generalized
central charge \agauge, where the Lagrange multipliers $\lambda_G$, $\lambda^\alpha$,
are related to the gauge and superpotential couplings via \rellamal, \glamNAbel, while
$d^a$, defined in \JgAbel, are functions on coupling space given by the solution of
\Gaugemax. One can think of \agauge\ as providing a potential on the space of theories.
Fixed points of the RG correspond to stationary points of this potential; RG trajectories are
governed by the gradient flow equation \betaamax, and its analog for gauge interactions.

We applied our results to the study of the vicinity of fixed points of the RG, rederiving and
extending the conditions for exact marginality of couplings and the behavior of anomalous
dimensions found in \GreenDA. We presented evidence that the metric on coupling space
associated with the flow \betaamax\ is the Zamolodchikov metric \metricG. Since this metric
is positive definite, the central charge $a$ decreases along RG flows. This generalizes
the results of \refs{\KutasovUX,\BarnesJJ} to a wide class of theories, including those
without a Lagrangian description.

Our discussion helps clarify the role of accidental symmetries in
supersymmetric RG flows. Such symmetries are associated with real superfields
$J_a$, whose dimension in the SCFT $\P$ is larger than two, but which can mix
with the supercurrent multiplet (as in \JgAbel) somewhere along the RG flow.
To understand their appearance it is useful to think about the space of theories
as a manifold, with coordinates given by the couplings $(\alpha, g^\gamma)$ or,
equivalently, $(\lambda_G, \lambda^\gamma)$. Typically, we study this manifold
by specifying a UV fixed point (in this case, the SCFT $\P$), and perturbations,
\eg\ the superpotential \Wsuper. RG flows describe trajectories on this manifold.

It may happen that the above description near $\P$ develops a coordinate
singularity at a finite distance along an RG flow. To describe the full
space of theories we then need to move to a different coordinate patch,
which is regular there. A natural candidate for such a patch is the vicinity of
the IR fixed point of the RG flow. We can repeat the discussion of this paper
around that fixed point, and then match the UV and IR descriptions in the region
in coupling space where both are valid.

As mentioned above, an example where accidental symmetries are important is
$\N=1$ SQCD with $N_f<{3\over2}N_c$, where the IR fixed point is the (IR free)
magnetic gauge theory of \SeibergPQ. The singularity of the
UV coordinates on theory space (the gauge coupling of the asymptotically
free electric theory) is in this case believed to be associated with the pole
in the NSVZ $\beta$-function \betal. The magnetic theory can be viewed as
providing a description of the patch of theory space that lies beyond the
region of validity of the electric description. The fact that the regions
in theory space covered by the electric and magnetic variables overlap
is essentially the statement that the conformal window $({3\over2}N_c<N_f<3N_c)$
exists.\foot{Which also leads to an overlapping region of validity outside
this range.}

This example, and the fact that in general one expects every point in the space of
theories to be in the regime of validity of {\it some} description of the sort presented
in this note, with $\P$ either the UV or the IR fixed point, suggests that accidental
symmetries do not change our qualitative results, including the gradient flow property
\betaamax\ and the monotonicity of $a$ along RG trajectories.

Some of the issues that we discussed can be illustrated in an example, $\N=1$ SYM
with gauge group $SU(N_c)$ coupled to chiral superfields $X$ in the adjoint representation,
and $Q_i$, $\tilde Q^i$, $i=1,\cdots, N_f$, in the fundamental representation. This
theory was studied using $a$-maximization in \refs{\IntriligatorJJ,\KutasovIY,\KutasovUX},
and exhibits many interesting features. It is asymptotically free for $N_f<2N_c$, and
the IR dimensions of the adjoint and fundamental chiral superfields can be calculated
for all $N_f$ in this range.

In the limit $N_f,N_c\to\infty$ these dimensions depend on the continuous variable $x=N_c/N_f$.
As $x$ increases, the dimension of the adjoint field $X$ at the IR fixed point decreases;
at large $x$, it goes to zero like $1/x$. This means that the operators ${\rm Tr} X^{k+1}$
with $k>2$, which are irrelevant at the UV fixed point of the $SU(N_c)$ gauge theory, become
relevant at the IR fixed point, for sufficiently large $x$. This is  related to the fact that the
IR theory becomes more strongly coupled as $x$ increases. Identifying the SCFT $\P$ in our general
discussion with the IR fixed point of the gauge theory with $W=0$, gives, in general, a SCFT
with a large class of relevant perturbations, which can be used to deform the theory, as in
\Wsuper. The perturbations ${\rm Tr} X^{k+1}$, in particular, were seen in
\refs{\KutasovVE\KutasovNP-\KutasovSS} and subsequent work to give rise to a rich set of dynamical
phenomena, such as confinement, chiral symmetry breaking, generalized Seiberg duality etc.

The non-R global symmetry group at $\P$  includes in this case the symmetry that assigns charge
$B_X$ to $X$ and $B_Q$ to $Q$, $\tilde Q$. The condition that this symmetry is non-anomalous
and thus conserved at $\P$ is $N_fB_Q+N_C B_X=0$.  In addition, there
are in general some accidental symmetries, which were discussed in \KutasovIY.

By tuning $N_f$, $N_c$, $k$, one can make the perturbation
\eqn\ggkk{W=g_k {\rm Tr} X^{k+1}}
slightly relevant. This gives an example where one can use the $\epsilon$-expansion discussed
around eq. \ireps. Of course, one can also obtain exact results for the scaling dimensions
at the IR fixed point of \ggkk, which agree with the $\epsilon$-expansion by construction.

One can also use this class of theories to study flows in which the UV fixed point does not have
any non-R symmetries. One way to do that is to take the SCFT $\P$ in the discussion of the previous
sections to be the IR fixed point of the RG flow \ggkk. The resulting theory has a unique R-symmetry,
under which $X$ has R-charge $2/(k+1)$ \KutasovNP. This symmetry belongs to the superconformal
multiplet; the corresponding current was denoted by $J^\mu_\P$ in \JgAbel.

Further perturbing the theory by adding a (relevant) superpotential $\delta W=g_{k'} {\rm Tr} X^{k'+1}$,
with $k'<k$, leads to a flow of the sort considered in \Wsuper. Since in this case there are no currents
that can mix with $J_\P$, the generalized central charge is given by \alambda\  with $d^a=0$ and
$R_\P=2(k'+1)/(k+1)$. Thus, it looks like $a$ decreases without bound as the Lagrange multiplier
$\lambda$ increases. This cannot continue indefinitely, for example because $a$ is known
to remain positive in this system all the way to the IR. In fact, what happens is that at some finite
$\lambda$ an accidental symmetry appears, which mixes with the current $J_\P$ inherited from
the UV theory, and at long distances the $U(1)_R$ in the current multiplet approaches the one
under which $X$ has R-charge $2/(k'+1)$.

This can be understood by embedding the above RG flow into one that includes as its
UV fixed point the IR fixed point of the gauge theory with vanishing superpotential described above.
As we mentioned, that theory has a non-R symmetry under which the adjoint field $X$ has an
arbitrary charge; hence it contains  the symmetry that becomes part of the superconformal
group in the IR limit of \ggkk, as well as its analog with $k\to k'$. Thus, in this larger theory we
can track how the symmetry changes from one to the other when we turn on $\delta W$ above
(see section 4 of \KutasovUX).

We expect this example to be representative of what happens in general in RG flows where the
UV fixed point has no global currents $J_a$ that can mix with the $U(1)_R$ current $J_\P$. Naively,
the central charge \aNAbelc\ decreases without bound in these cases, but at some finite value
of the couplings $\lambda^\alpha$ accidental symmetries must alter that conclusion. As mentioned
above, to obtain a full description of such RG flows one has to match the descriptions
around the UV and IR fixed points. We expect the qualitative properties of these RG flows
to be the same as those discussed above.

There are a number of possible directions for further work  along the lines of our construction.
Although the Lagrange multiplier technique seems to be very useful in clarifying the pattern of
RG flows for a wide class of theories, it remains somewhat mysterious. In particular, it would
be nice to have a more intrinsic definition of the generalized central charge $a$ along RG flows,
\eg\ in terms of correlation functions  of local operators, or some other QFT data.

A related issue is the metric on theory space that enters the gradient flow \betaamax.
We saw that near fixed points, this metric is given to leading order by the Zamoldochikov metric
\metricG, and suggested that this may be true more generally. It would be nice to understand
whether this is indeed the case. The metric in \betaamax\ is also important for the coordinate
singularities associated with the breakdown of the UV variables discussed above. It would be
nice to understand these singularities better.

While our construction provides a strong argument for the $a$-theorem in a wide class
of $\N=1$ QFT's, there are important classes of deformations that are not covered by it. One
class corresponds to turning on expectation values of local operators, which take one along
moduli spaces of SUSY vacua. Many theories, including those mentioned earlier in this section,
are known to have such moduli spaces, and it would be interesting to generalize our arguments
to such spaces (see \BarnesJJ\ for some comments on this issue). Another class corresponds to
turning on Fayet-Iliopoulos terms for $U(1)$ factors in a gauge theory.

A counter-example to the $a$-theorem was proposed in \ShapereUN. It would be 
interesting to understand it in the framework of our construction, and in particular to determine 
whether it is consistent with our results. This may require a better understanding of moduli 
spaces of vacua of the sort mentioned in the previous paragraph.

\bigskip

\noindent{\bf Acknowledgements}: We thank A. Schwimmer for
discussions. This work is supported in part by DOE grant
DE-FG02-90ER40560, the National Science Foundation under Grant
0529954, and the BSF -- American-Israel Bi-National Science
Foundation. DK thanks the Weizmann Institute for hospitality during
part of this work.

\listrefs
\end